 \documentclass[final,5p,times,twocolumn,authoryear]{elsarticle}

\usepackage{amssymb,hyperref}
\usepackage{lipsum}

\journal{Astronomy $\&$ Computing}

\begin{document}

\begin{frontmatter}



\title{Using GMM in Open Cluster Membership: An Insight}


\author[first]{Md Mahmudunnobe}
\affiliation[first]{organization={Wayne State University},
            addressline={42 W Warren Ave}, 
            city={ Detroit},
            postcode={48202},              state={MI},
            country={USA}}
\author[second, third]{Priya Hasan}
\affiliation[second]{organization={Maulana Azad National Urdu University},
            addressline={Gachibowli}, 
            city={Hyderabad},
            postcode={500 032},              state={Telangana},
            country={India}} 
 \affiliation[third]{organization={Inter-University Centre for Astronomy and Astrophysics},
            addressline={Post Bag 4
Ganeshkhind,
Savitribai Phule Pune University Campus}, 
            city={Pune},
            postcode={411 007},              state={Maharashtra},
            country={India}} 
\author[second]{Mudasir Raja}
\author[second]{Md Saifuddin}
\author[second]{S N Hasan}

\begin{abstract}
The unprecedented precision of Gaia has led to a paradigm shift in membership determination of open clusters where a variety of machine learning (ML) models can be employed. In this paper, we apply the unsupervised Gaussian Mixture Model (GMM) to a sample of thirteen clusters with varying ages ($log \ t \approx$ 6.38-9.64) and distances (441-5183 pc) from Gaia DR3 data to determine membership.  We use ASteca to determine parameters for the clusters from our revised membership data. We define a quantifiable metric Modified Silhouette Score (MSS) to evaluate its performance. We study the dependence of MSS on age, distance, extinction, galactic latitude and longitude, and other parameters to find the particular cases when GMM seems to be more efficient than other methods. We compared GMM for nine clusters with varying ages but we did not find any significant differences between GMM performance for younger and older clusters. But we found a moderate correlation between GMM performance and the cluster distance, where GMM works better for closer clusters.  We find that GMM does not work very well for clusters at distances larger than 3~kpc.
\end{abstract}




\end{frontmatter}




\section{Introduction}
\label{introduction}

The traditional method for finding and identifying open star clusters involved looking for star overdensities in the sky, supplemented with color-magnitude diagrams and/or proper motion or spectroscopic data. 










The \cite{dias2002new} catalog is a compilation of heterogeneous data from various observations and methods to determine cluster parameters. Determination of cluster parameters requires not only a homogeneous set of data, but also similar techniques of analysis. The homogeneous catalog by \citep{kharchenko2013global} used data from USNO CCD Astrograph Catalog (UCAC) and PPMXL \citep{zacharias2013fourth, roeser2010ppmxl}. The Two Micron All Sky Survey (2MASS) data has also been widely used to study star clusters \citep{skrutskie1997two, skrutskie2006two}.  
Gaia is a homogeneous catalog of all-sky data with  unprecedented astrometric precision and hence makes it ideal to identify members in existing open clusters (OCs) and  dismiss several asterisms.
It also makes it possible to identify subgroups and structures in the regions surrounding OCs. 

Machine Learning has been increasingly used in astronomy to  analyse large amounts of data.
  Supervised machine learning algorithms are algorithms that are used to learn a relationship between a set of measurements and a target variable by making use of labelled data \citep{mahabal2008automated,brescia2012detection,ishida2019optimizing}. This method  is used to predict the value of the target variable. 
  In supervised learning techniques, the model parameters are estimated  from the data, and these estimations from the training set help define the model. The model is applied on the data to generate accurate predictions. However, in the case of errors or biases in the training set,  this method  can lead to wrong results.
  
  Unsupervised learning refers to the process of teaching a machine to do a task using data that has not been categorised or labelled in any way, and then letting the machine's algorithm  make decisions independently based on the results of those calculations. These include clustering analysis, dimensionality reduction, visualisation, and the identification of outliers. Tools of this kind are of extreme importance in the field of scientific study, especially where no prior assumptions are made. This is due to the fact that they may be utilised to either produce new discoveries or extract new knowledge from datasets, without using a training set that could be biased or with errors.

\cite{gao20205d} employed Principal Component Analysis in addition to a Gaussian Mixture Model (GMM) in order to identify members and characterize the tail of the cluster NGC~2506. \cite{bhattacharya2021tidal} identified the tails of NGC~752  and \cite{agarwal2021ml} introduced the membership determination method ML-MOC for a sample of open clusters.

\cite{2020ApJ...894...48G} employed Principal Component Analysis and GMM  to identify extra-tidal stars.and found 2301 stars closely related to the cluster, 147 of which are likely extra-tidal stars. In an earlier paper \citep{Mahmudunnobe_2021}, we used the supervised technique of Random Forest to find membership of stars in a sample of nine clusters. To avoid the dependence of labelled data in supervised techniques, in this paper, we applied the unsupervised clustering technique GMM  on a sample of thirteen clusters with Gaia DR3 data \citep{2023A&A...674A...1Gca} and use it to determine  membership of stars at the low mass end and derive parameters for our clusters. Our cluster sample has a wide range of ages and distances.  We define a quantifiable metric Modified Silhouette Score (MSS) to evaluate its performance and compare its value for our sample. We compare the spectroscopic data of members identified by 
\citep{cantat2018gaia} and this work using APOGEE and GALAH data to validate our member sample. We use ASteca to determine parameters for the clusters from our revised membership data.

\section{Cluster Sample}
The basic parameters of the 13 selected clusters are given in Table \ref{clusdatg} which shows the coordinates of these clusters $\alpha$ and $\delta$,  the radius that contains half the number of members from the same reference $r50$, the logarithm of age $log \ t$, the distance to the cluster in parsecs $d$ and the galactocentric distance in parsecs $GC$  from  \citep{cantat2020clusters}. The sample covers a large range of ages ($log \ t \approx$ 6.38-9.63) and distances (441-5183 pc). 

\begin{table}[h]
\tiny
\begin{tabular}{lllllllll}
\hline

Cluster &   $\alpha$ & $\delta$ & $l$ & $b$ &$r50$ &  $log\ t$  & $d$ &  $GC$  \\  
   &   (deg) & (deg) &(deg) & (deg)  & (pc)  &  & (pc)& (pc)   \\ 

\hline
\hline
\\

NGC~752    & 29.22  & 37.79 & 136.9 & -23.3 &0.049 & 9.18   & 441 & 8640 \\
IC 4651        & 261.21  & -49.92 & 340.1& -7.9 & 0.23 & 9.32 &  920 & 7488 \\
NGC 2539       & 122.66  & -12.83 & 233.7 & 11.1 &0.19 & 8.83 &  1243 & 9137 \\
NGC 2099        & 88.07  & 32.54 &177.6 & 3.1 & 0.16 & 8.78 &  1299 & 9775 \\
NGC 581        & 23.34  & 60.66 & 128.0 & -1.8 & 0.062& 7.44 &  2502 & 10075 \\
NGC 6823       & 285.79 & 23.32 & 59.4 & -0.1 &0.074& 6.38  & 2330 & 7430 \\
NGC 2243       & 97.4   & -31.28& 239.5 & -18.0&0.046 & 9.64  &3719 & 10584 \\
IC 1805        & 38.21  & 61.47 & 134.7 & 0.9 &0.11 & 6.88  & 1964 & 9821 \\
NGC 7142        & 326.29  & 65.78 & 105.4 & 9.5 &0.10 & 9.55  & 2040 & 9241 \\
NGC 6791 & 290.22 & 37.78 & 69.9 & 10.9 & 0.068 & 9.8  & 4231 & 7942\\
NGC 2141       & 90.73  & 10.45 &198.0 & -5.8& 0.073 & 9.27 &  5183& 13339 \\
NGC 1893       & 80.72  & 33.44 &173.5 & -1.6 & 0.085& 6.64 &  3222 & 11546 \\
NGC~2682& 132.85 & 11.8 & 215.7 & 31.9   &0.166 & 9.63   &  899 & 8964 \\
\hline

\end{tabular}
\caption[Basic cluster parameters]{Basic cluster parameters\\ \citep{cantat2020clusters}}
\label{clusdatg}
\end{table}

\section{The GMM Method}


The GMM is a parametric machine learning model and is based on the assumption that the data is a combination of two or more Gaussian distributions.

In the case of star clusters,  we have two distinct groups of stars: members and field stars. We can assume that the members follow a normal distribution in the feature space made up of ($\alpha$, $\delta$, $\pi,\ \mu_{\alpha}$, and $\mu_{\delta}$), and are clustered.
However, the field stars would be distributed in a random manner and will not follow a broader normal distribution.

When we have a large sample area in comparison to the cluster size, previous researchers \citep{gao2018memberships} and \citep{agarwal2021ml} found that GMM does not work very well.
This is because in the sample, the field stars dominated and the cluster stars were difficult to be identified.

It was pointed out by \citep{cabrera1990non,1979VatOP...1..283D} that when it comes to cluster membership, one strategy that should be employed in models like GMM, is by making sure that the following conditions are met:
\begin{itemize}
    \item 
    The ratio of the number of member stars to the number of field stars in the sample area should be high. This will make sure that in the dataset the member group is the primary group of stars.

    \item 
    There should be a difference between the peak positions of the field star distributions and the member star distributions in feature space. If such is not the case, then the GMM model will not be able to differentiate between these two groups of stars.
\end{itemize}

With reference to the second criterion, it is highly improbable that the peak of the member star distribution and the field star distribution will coincide in feature space. But if it does take place, then we will have to try to limit the size of our sample region in order to cut down on the number of field stars. 

Keeping this in mind,  we query the data using a cone search for a specific position in the sky (i.e., $\alpha$ and $\delta$ ) and a search radius, \textit{r}. Then we extract all the stars that lie around that position within the given search radius. A larger radius will increase the number of field stars and violate the first condition. For a smaller search radius, the field stars, and the member stars both have the highest number density in the center, i.e., the peak of their distribution in $\alpha$ and $\delta$ overlaps, and breaks the second condition. Hence, it is better not to use $\alpha$ and $\delta$ as one of our input features in GMM. Instead, we can use a smaller search radius to ensure that all the stars in the dataset are close to the cluster center. 

The method to ensure that the first condition is met is by applying GMM in a more constrained region and using an optimal range of all feature variables. If our operational radius is too large, then we will have a significant number of field stars and the member group will no longer function as the primary group. If our range is too confined, then we will not have many field stars and GMM will attempt to locate two distinct groups within the members. To ensure this condition, we extract stars from a certain region using a distance cutoff that we will explore in the following sections. The Gaia data was filtered through quality checks: parallax $>$ 0, parallax/error $> 3$, errors in $\mu_{\alpha}$ and $\mu_{\delta} < 0.3$. This has been done to ensure that the data used is reliable and not noisy.

\section{The Performance Metric: Modified Silhouette Score (MSS)}
In this section, we shall introduce the Modified Silhouette Score (MSS) that we  defined for GMM.  

The Silhouette Score (SS) is a metric used to calculate the performance of a given clustering technique and validate the clustering algorithm. When using the silhouette approach, each point's silhouette coefficient is calculated, which indicates how well the point belongs to a cluster rather than to some other cluster. It gives a graphical illustration of how accurately each cluster has been identified.

The SS is a representation of how far an object is from other clusters in comparison to objects in its own cluster. A high value implies the object is well-matched to its own cluster and poorly matched to neighboring clusters. The value of the SS lies in the interval $[1, -1]$. The goodness of a clustering technique is indicated if the majority of the objects have high values of SS. The silhouette coefficient for the sample point is defined as,

$$s = \frac{b-a}{\max(a,b)}$$
\noindent
where $a$ is the mean distance between a sample and all other points in the same class and $b$ is the mean distance between the sample and all other points in the next class.

The problem in this metric is that the silhouette score assumes that all clusters are dense and well separated, which is not true in the case of star clusters. We have a member set, that is compact and dense in all the feature-variable spaces. But our field stars are random and uniform in all feature spaces and is mixed with the member set in feature space.  
Even when we have a  good separation of member and field stars, for a member star the value of $b$ would be close to the value of $a$, as the field stars are uniformly found all around the member set. Hence, the silhouette score will be close to 0. 

We shall use another property of our sample in this case. The standard deviation ($\sigma$) of the members would be small due to the cluster's compact nature. On the other hand, the field stars are dispersed evenly, which means that their $\sigma$ ought to be high. Therefore, in our case, $\sigma$ as a metric may be more helpful than the comparison of the distance between clusters. In light of this, a new metric is proposed for evaluating performance for clustering by an unsupervised model by making certain adjustments to the silhouette score. This newly proposed metric is named as the Modified Silhouette Score (MSS) and is denoted by

\begin{equation}
    MSS = \frac{1}{k} \sum_{i=1}^{k} \frac{(\sigma_{i, field} - \sigma_{i, member})}{\max(\sigma_{i, field}, \sigma_{i, member})}
    \label{eq:mss}
\end{equation}
\noindent
where $k$ is the total number of features and  $\sigma_{i, field}$ and $\sigma_{i, member}$ denote the $\sigma$ of the feature $i$ for field stars and members  respectively.

We would expect that a well-performed model would show the members to be distributed normally with a very small $\sigma$ and the field stars to be uniformly distributed, i.e., with a high $\sigma$.

In this case, we would have $\sigma_{field} >> \sigma_{member}$, therefore the numerator will be $\sigma_{ field} - \sigma_{i, member} \approx \sigma_{i, field}$. This will result in an MSS value very close to 1. On the other hand, for a poor performance model, the member and field stars both will have a similar random distributions. Thus, the numerator will be close to 0, resulting in an MSS value around 0. One special case is when the predicted field star group shows a stronger normal distribution (thus having low $\sigma$), but the predicted member group is distributed randomly (a larger $\sigma$). In this case, the numerator will be $\sigma_{field} - \sigma_{i, member} \approx -\sigma_{i, member}$ and the MSS value will be around -1. So, a strong negative MSS value will likely indicate that the model was able to distinguish well between member and field stars, but it mislabeled the groups. The predicted member group is the field star group and vice-versa.

\subsection{Performance of GMM in Simulated Data}
We used a simulated dataset of normally distributed members and randomly distributed field stars to check how changing the range of the features influences the performance of GMM. We have two features in each simulation: \textit{feature~1} and \textit{feature~2}, analogous to $\mu_{\alpha}$ and  $\mu_{\delta}$.  We denote the \textit{half-width} for a given variable by \textit{hw}. If the chosen center of the variable is $x$, then we take $x \pm hw$ values: the chosen range for the variable is [$x-hw, x+hw$]. 

\begin{figure}[h]
\centering 	\includegraphics[width=\columnwidth]{ 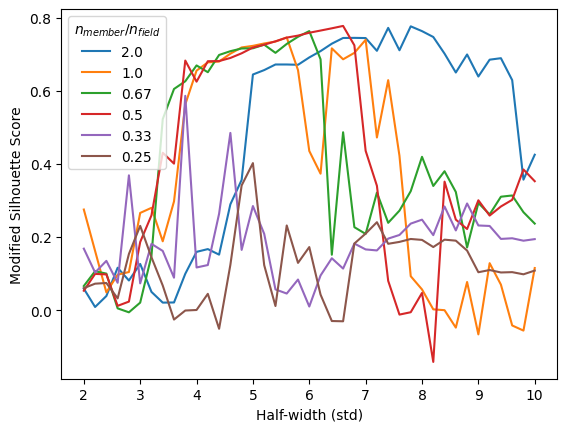}
    \caption[MSS vs Half-width]{MSS vs Half-width for the varying ratio of member and field stars}
\label{fig:mss_vs_hw}
\end{figure}

We ran the simulations for 40 different values of $hw$ varying between $2 \sigma$ to $10 \sigma$. Each of the simulations was run 20 times ($n_{trial}$ = 20) and we took the average value of the MSS metric. Then we changed the number of field stars ($n_{field}$) while keeping the grid size and the number of members constant. This was to study the effect of the field star density on GMM. For each value of $n_{field}$, we ran the system again for 40 different values of \textit{hw} and with 20 iterations each time. The results of the simulations are shown in  Fig. \ref{fig:mss_vs_hw}. We note a common trend for all values of $n_{field}$. Each of them has a low MSS value at first. Then in the middle as $hw$ increases to around $4 \sigma$ to $6 \sigma$, there is a flat peak at MSS. Then for a larger value of  $hw$, the MSS decreases. For a smaller half-width, there are only a very few field stars and GMM tries to separate the member stars into two different groups. For a larger cutoff, GMM fails to capture the prominent distribution. Only for an optimal range of $hw$ values, GMM works well.

Another point to note from Fig \ref{fig:mss_vs_hw} is the influence of the field star density. As the grid size is constant, the field star density increases with increasing values of $n_{field}$, i.e., decreasing value of $n_{member}/n_{field}$. We can see that for a smaller field star density, we get a good MSS for a relatively high value of $hw$ ($5.5 \sigma$) and it stays good up to a relatively high $hw$ value ($7.5\sigma$). This trend is observed as we need to increase the range of the variable to get enough field stars in our dataset due to the small field star density. As the ratio of the members to field stars increases, the optimum region shifts towards the lower values of $hw$ (around $3.3\sigma$ to $5.5\sigma$). Then, once the ratio of the members to field stars decreases ($n_{member}/n_{field} = 0.25$), we have a low MSS. The maximum value of the MSS metric is still low for high field star densities ($< 0.4$).

From this analysis, we infer that the quality of the GMM model depends on choosing an optimal filter/cutoff for the features. In our analysis, we derived this optimal cutoff for parallax and proper motion empirically. In that case, we get good MSS values.

\section{GMM Analysis}
As we discussed earlier, for a smaller search radius, $\alpha$ and $\delta$ can sometimes affect the performance of GMM, as both groups of stars often have a similar peak position and similar distribution. Fortunately, for a smaller radius, all the stars are very close to the cluster center, so it is not necessary to include them in our model. Thus, we ran the  analysis with only three important variables of choice 
$pmra\ (\mu_{\alpha})$\footnote{$pmra=\mu_{\alpha}$ is actually $\mu_{\alpha}^*=\mu_{\alpha} \times cos \delta$. Henceforth, it will be referred to as $pmra$}, $pmdec\  (\mu_{\delta})$ and  parallax~($\varpi$). Further, we need to get rid of noisy data, by applying quality filters.  We used the following filters in our analysis: errors in $\mu_{\alpha}$ and $\mu_{\delta}$ to be less than 1 and $\varpi/{errors} > 3$. Then we normalized each of the features before feeding it to our model.

We ran the 2-component GMM algorithm with 5 different initial conditions and chose the best one (by setting $n_{init}$ = 5). This is because the convergence of the Expectation-Maximum algorithm, used to train the GMM model, is only guaranteed to a local optimum, not to the global optimum.  We have used the default type “full” as we did not want to make any assumptions.

For the cutoff in the parameter space,  \cite{gao2018memberships, agarwal2021ml} used a trial and error method to choose the cutoff. In our analysis, as a first filter, we only took stars with proper motions ($pmra$, $pmdec$) between -20 to 20 $mas$/yr. The reasoning is that any star with higher than 20 $mas$/yr proper motion will always escape the cluster. 

\begin{figure}
\centering    \includegraphics[width=0.9\columnwidth]{ 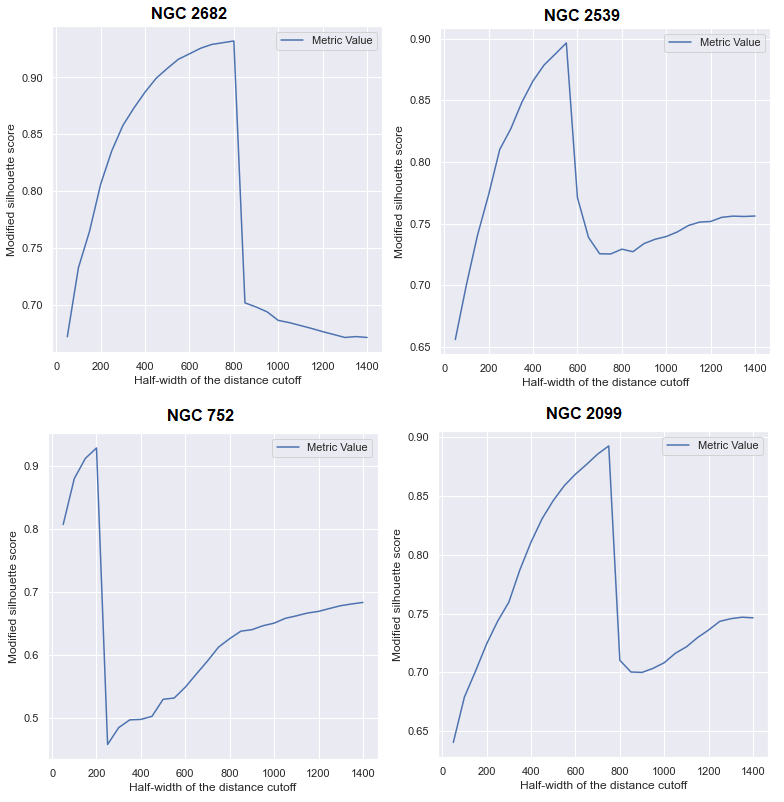}
    \caption{MSS value vs distance cutoff (in pc) for four clusters of our sample. We can see that initially, MSS increases with the distance cutoff up to a certain point, and then the MSS drops significantly. This boundary point is taken as our optimal distance cutoff, which is found from the sharp drop-off point in the plots}
    
    \label{mss_hw_clusters}
\end{figure}

The cutoff for the distance (i.e., parallax) is not as straightforward as it depends on the cluster distance, its member density, and field star density along the line of sight. We used an empirical approach to choose the optimal distance cutoff. If the mean cluster distance (either from literature or by taking the mean of the star distances) is $d$ pc, then we change the distance filter from $d \pm 50$ pc up to $d \pm 1400$  pc, run the GMM model with filtered stars, and record the model performance using MSS. Finally, we chose the cutoff, where the model performs best. Figure \ref{mss_hw_clusters} shows the MSS vs distance cutoff for four of our sample clusters. We can see that initially, MSS increases with the distance cutoff up to a certain point, and then drops significantly. This boundary point is taken as our optimal distance cutoff.

GMM divides the stars into two groups. We defined the member group as the group with a lower $\sigma$ in the feature spaces. It also assigns a membership probability for each of the sample stars to be in the member group. We followed a similar approach to choose the optimal member threshold. We varied the member threshold, measured the MSS of the model, and finally chose the one with the highest MSS value\footnote{All the code for GMM is available at \url{https://github.com/mahmud-nobe/Cluster-Membership/tree/master/GMM}}.

\section{Revised Membership samples}
\begin{figure}
\centering  
\includegraphics[width= \columnwidth]{ 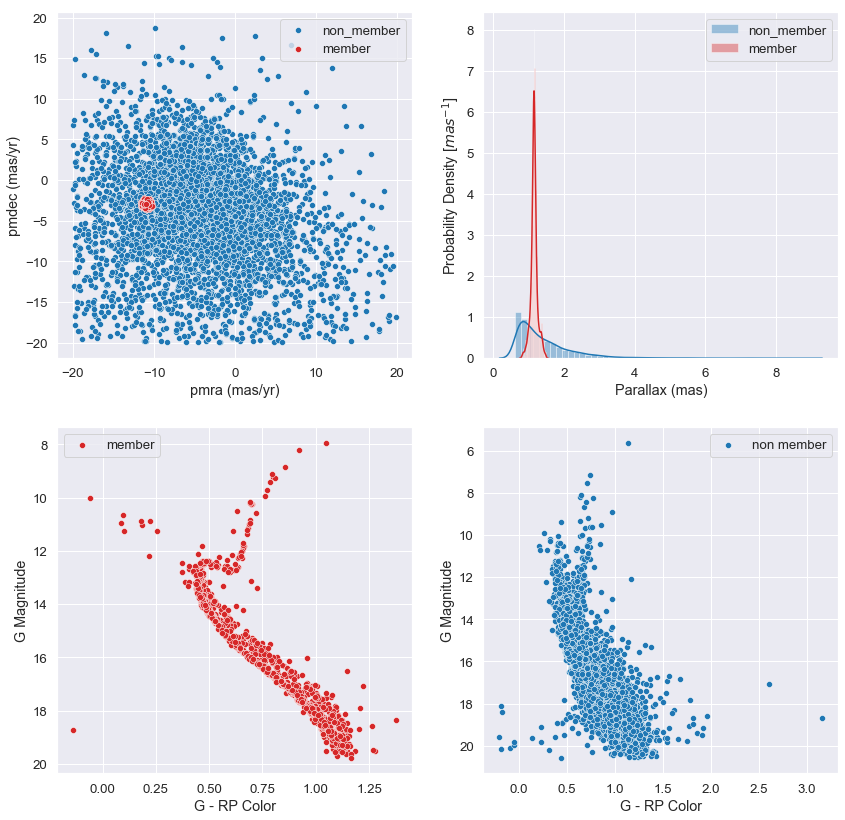}
\caption[GMM results for M67]{GMM results for M67 with distance cutoff of 750~pc and member threshold of $P_{memb} \geq 0.95$. The upper left shows the plot in $\mu_{\alpha}$ vs $\mu_{\delta}$, the upper right shows parallax, the lower left is the CMD for cluster stars and the lower right is for field stars.}
    \label{m67}
\end{figure}

\begin{figure}
\centering  
\includegraphics[width=\columnwidth]{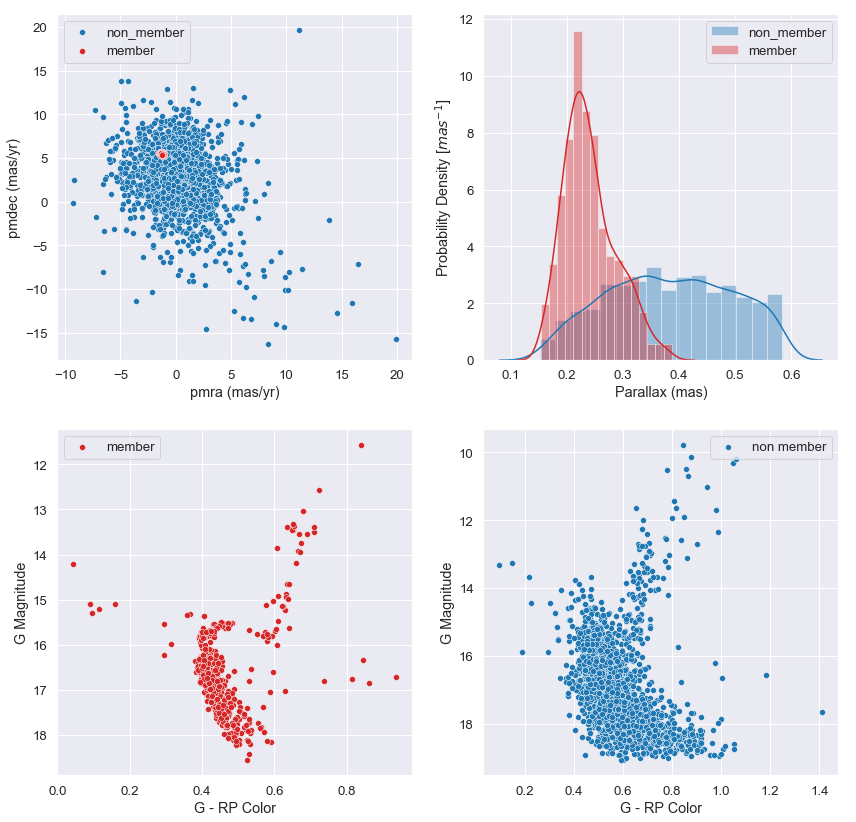}
\caption[GMM results for NGC~2243]{GMM results for NGC~2243 with distance cutoff of 2450 pc and member threshold of $P_{memb} \geq 0.95$ with the two groups in red (cluster)  and  blue (field) . The upper left shows the plot in $\mu_{\alpha}$ vs $\mu_{\delta}$, upper right shows parallax, lower left is the CMD for cluster stars and lower right is for field stars.}
\label{n2243}
\end{figure}

\begin{figure}
\centering  
\includegraphics[width=\columnwidth]{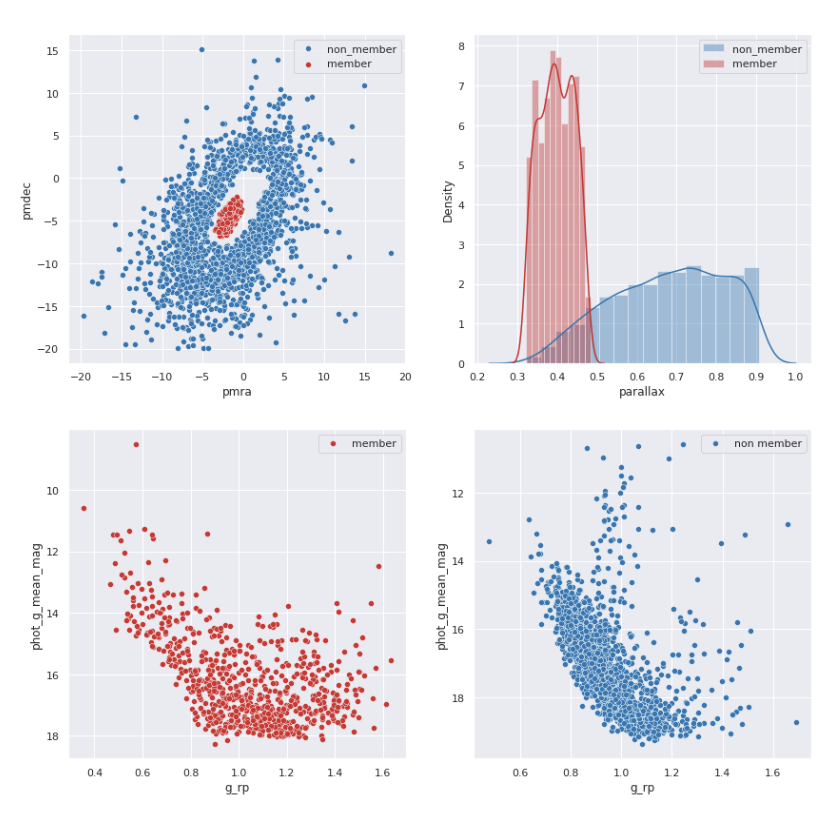}
\caption[GMM results for NGC~6823]{GMM results for NGC~6823 with distance cutoff of 900 pc and member threshold of Pmemb $\geq$ 0.95 with the two groups in red (cluster)  and  blue (field) . The upper left shows the plot in $\mu_{\alpha}$ vs $\mu_{\delta}$, upper right shows parallax, lower left is the CMD for cluster stars and lower right is for field stars.}
\label{n6823}
\end{figure}

\begin{figure}
\centering 
\includegraphics[width=\columnwidth]{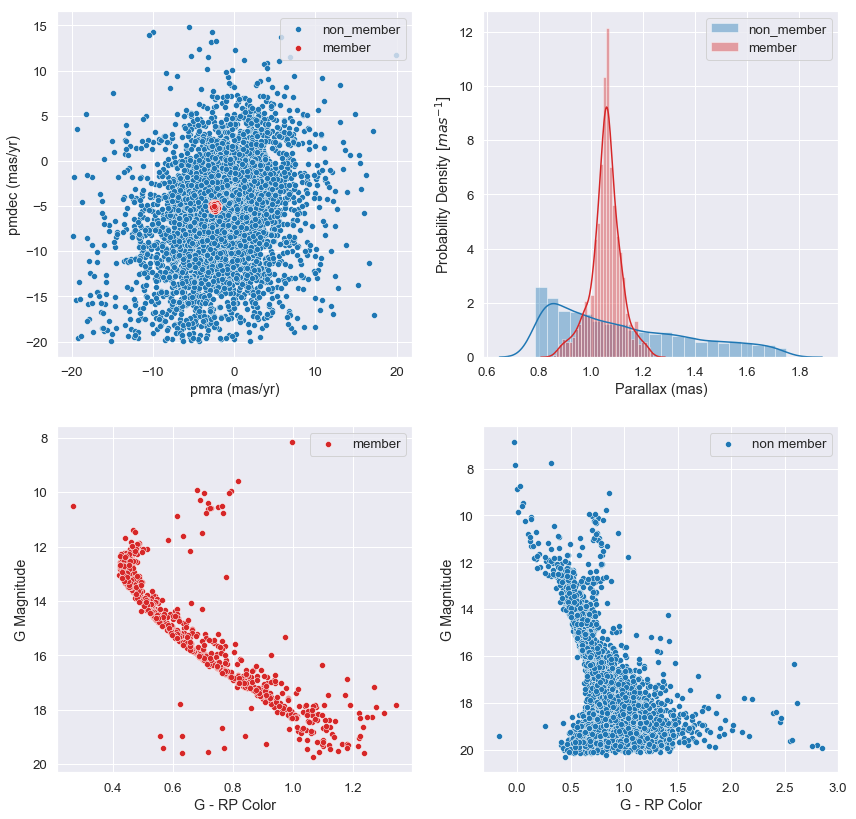}
\caption[GMM results for IC~4651]{GMM results for IC~4651 with distance cutoff of 350 pc and member threshold of $P_{memb} \geq 0.95$ with the two groups in red (cluster)  and  blue (field) . The upper left shows the plot in $\mu_{\alpha}$ vs $\mu_{\delta}$, upper right shows parallax, lower left is the CMD for cluster stars and lower right is for field stars.}
    \label{i4651}
\end{figure}

\begin{figure}
\centering  
\includegraphics[width=\columnwidth]{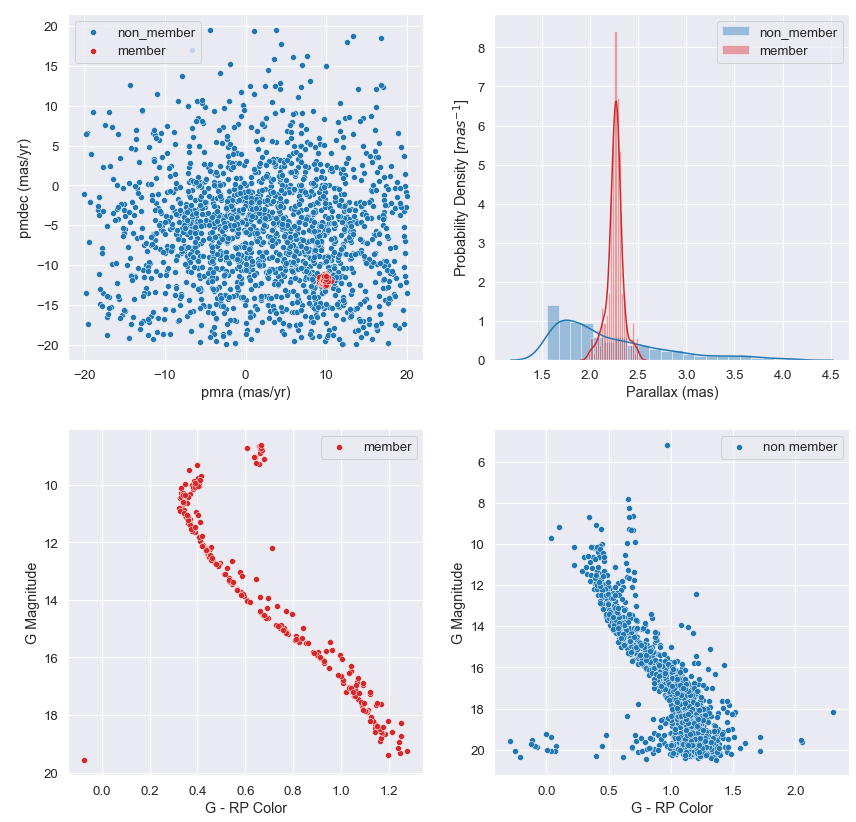}
\caption[GMM results for NGC~752]{GMM results for NGC~752 with distance cutoff of 200 pc and member threshold of $P_{memb} \geq 0.95$ with the two groups in red (cluster)  and  blue (field) . The upper left shows the plot in $\mu_{\alpha}$ vs $\mu_{\delta}$, upper right shows parallax, lower left is the CMD for cluster stars and lower right is for field stars.}
\label{n752}
\end{figure}

\begin{figure}
\includegraphics[width=\columnwidth]{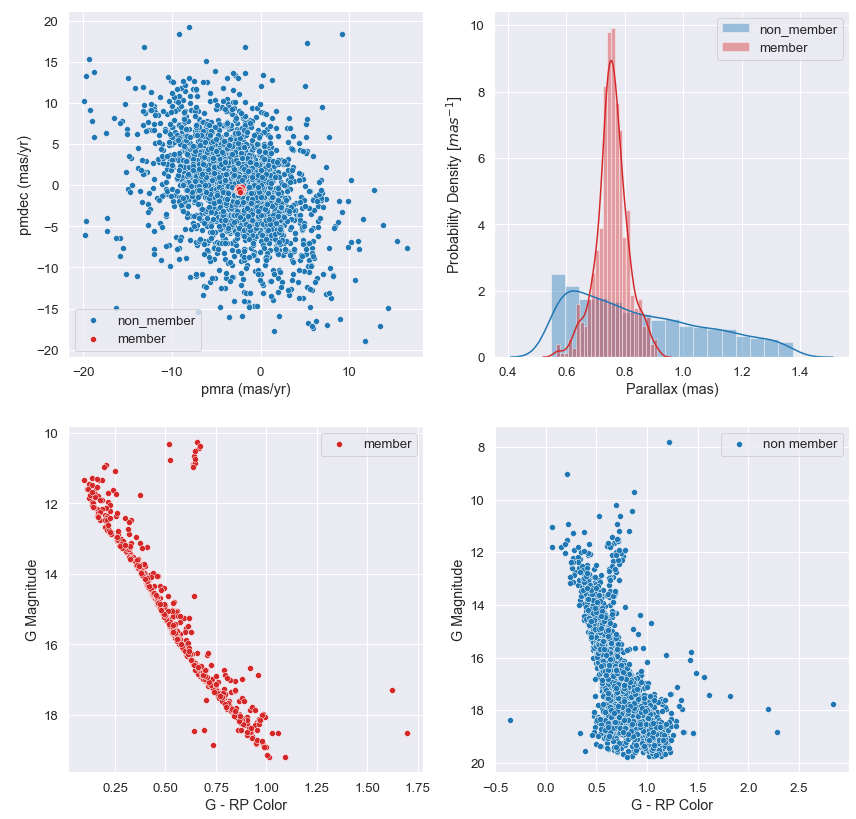}
\caption[GMM results for NGC~2539]{GMM results for NGC~2539 with distance cutoff of 550 pc and member threshold of $P_{memb} \geq 0.95$ with the two groups in red (cluster)  and  blue (field) . The upper left shows the plot in $\mu_{\alpha}$ vs $\mu_{\delta}$, upper right shows parallax, lower left is the CMD for cluster stars and lower right is for field stars.}
\label{n2539}
\end{figure}

\begin{figure}
\centering  
\includegraphics[width=\columnwidth]{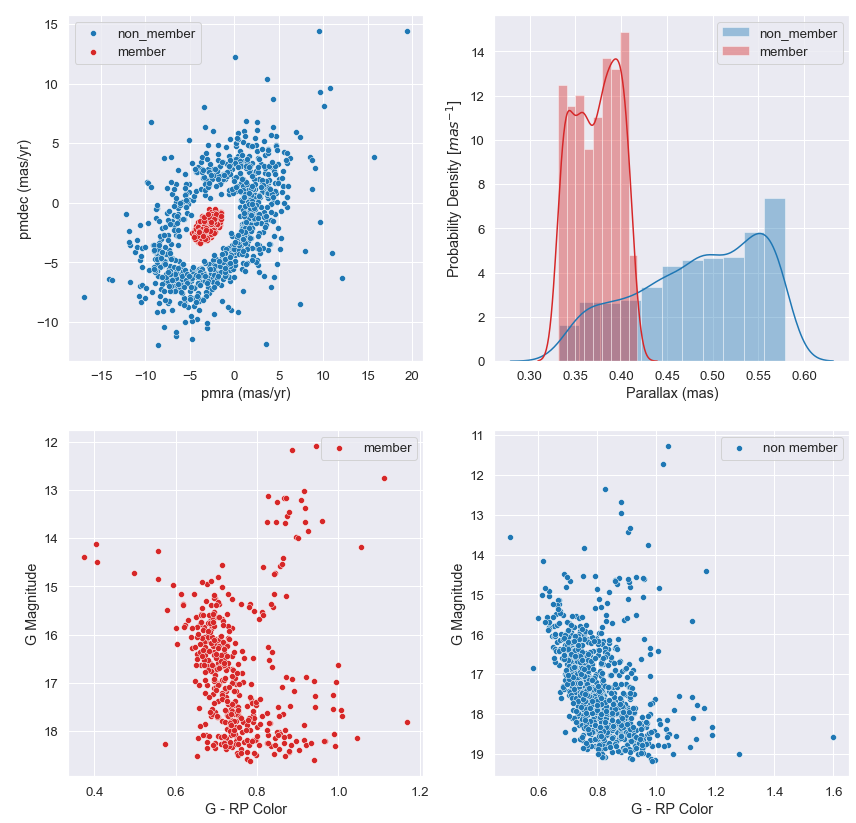}
\caption[GMM results for NGC~7142]{GMM results for NGC~7142 with distance cutoff of 650 pc and member threshold of $P_{memb} \geq 0.90$ with the two groups in red (cluster)  and  blue (field) . The upper left shows the plot in $\mu_{\alpha}$ vs $\mu_{\delta}$, upper right shows parallax, lower left is the CMD for cluster stars and lower right is for field stars.}
\label{n7142}
\end{figure}

\begin{figure}
\centering  
\includegraphics[width=\columnwidth]{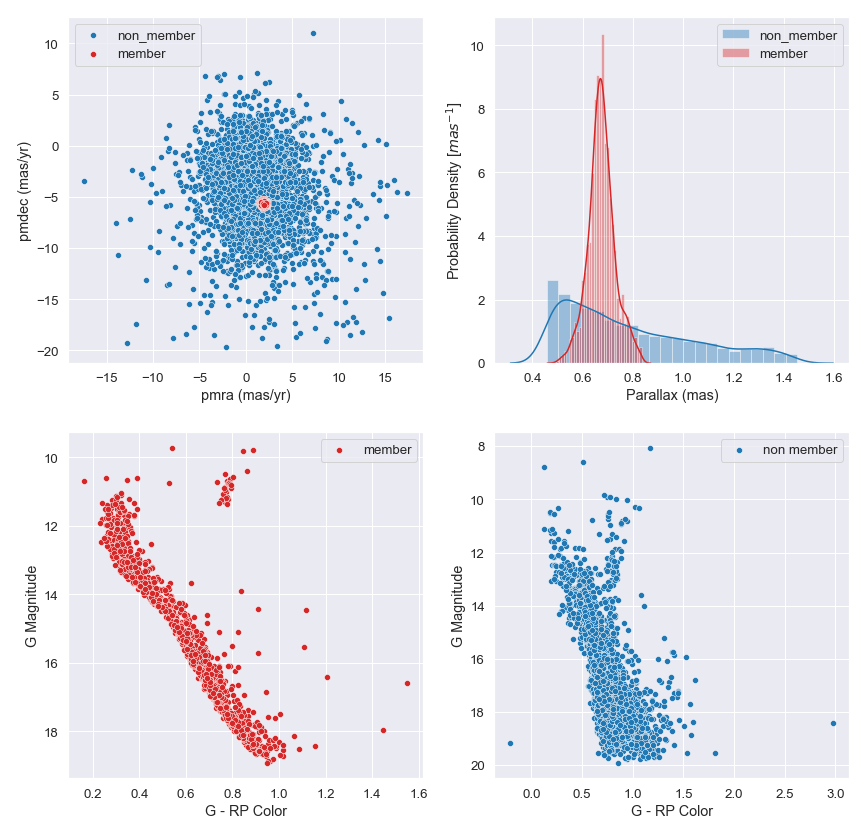}
\caption[GMM results for NGC~2099]{GMM results for NGC~2099 with distance cutoff of 750 pc and member threshold of $P_{memb} \geq 0.95$ with the two groups in red (cluster)  and  blue (field) . The upper left shows the plot in $\mu_{\alpha}$ vs $\mu_{\delta}$, upper right shows parallax, lower left is the CMD for cluster stars and lower right is for field stars.}
    \label{n2099}
\end{figure}

\begin{figure}
\centering  
\includegraphics[width=\columnwidth]{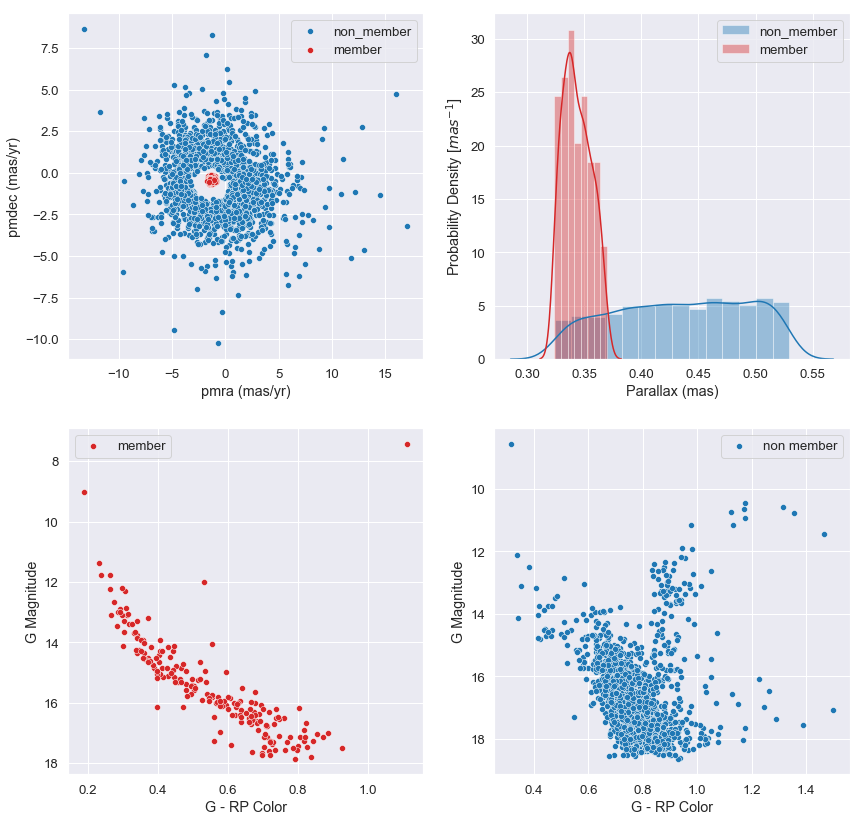}
\caption[GMM results for NGC~581]{GMM results for NGC~581 with distance cutoff of 600 pc and member threshold of $P_{memb} \geq 0.95$ with the two groups in red (cluster)  and  blue (field) . The upper left shows the plot in $\mu_{\alpha}$, upper right shows parallax, lower left is the CMD for cluster stars and lower right is for field stars.}
\label{n581}
\end{figure}

Figures \ref{m67} to \ref{n581} show the distribution of the two groups: cluster stars (blue) and field (red) for nine of our clusters. The upper left in each figure shows the plot in $\mu_{\alpha}$ vs $\mu_{\delta}$, the upper right shows parallax, lower left is the CMD for cluster stars and lower right is for field stars. The proper motion plots in the upper left, a very clear region occupied by members, is very compact and well defined. The parallax plots on the upper right show that the members and field stars have very clearly different peaks and the distribution of the parallax has smaller standard deviations for members. In the plots, non members were defined as stars with a membership probability $PMemb <= 0.2$ and members were stars with $PMemb >= 0.95$ for all clusters excluding NGC~7142 with $PMemb >= 0.90$. This was to ensure the confidence in our membership determination and avoid any field star contamination. In the case of some clusters  (Figs. \ref{n6823},  \ref{n7142},  \ref{n581}), there is a very clear separation of these stars, where the space between members and non-members is empty, i.e., devoid of stars with $PMemb$ between  0.2 and 0.95 indicating a very clear categorization of members  and non-members. In the case of the rest of our sample of clusters, this is not very well segregated  and there are stars with intermediate values of $PMemb$. 

\begin{figure}[h]
\centering  
\includegraphics[width=\columnwidth]{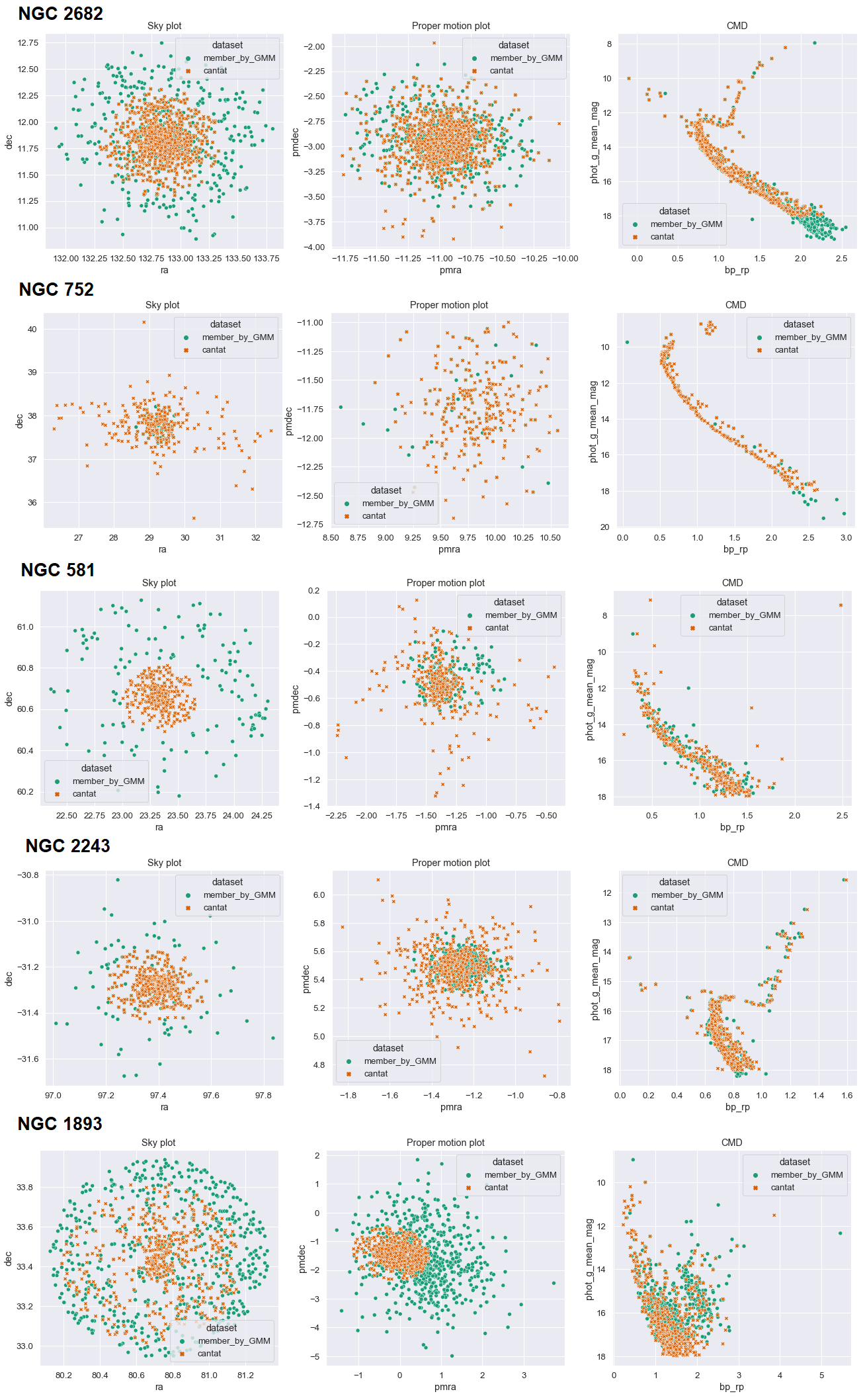}
\caption[Comparison of members]{Comparison of members defined by our method (green) and \cite{cantat2020clusters} (orange). The plots show pmra and pmdec in mas/yr, ra  and dec in degrees, $g_{mag}$ and bp-rp in mag.}
\label{cg1}
\end{figure}

\begin{figure}[h]
\centering  
\includegraphics[width=\columnwidth]{ 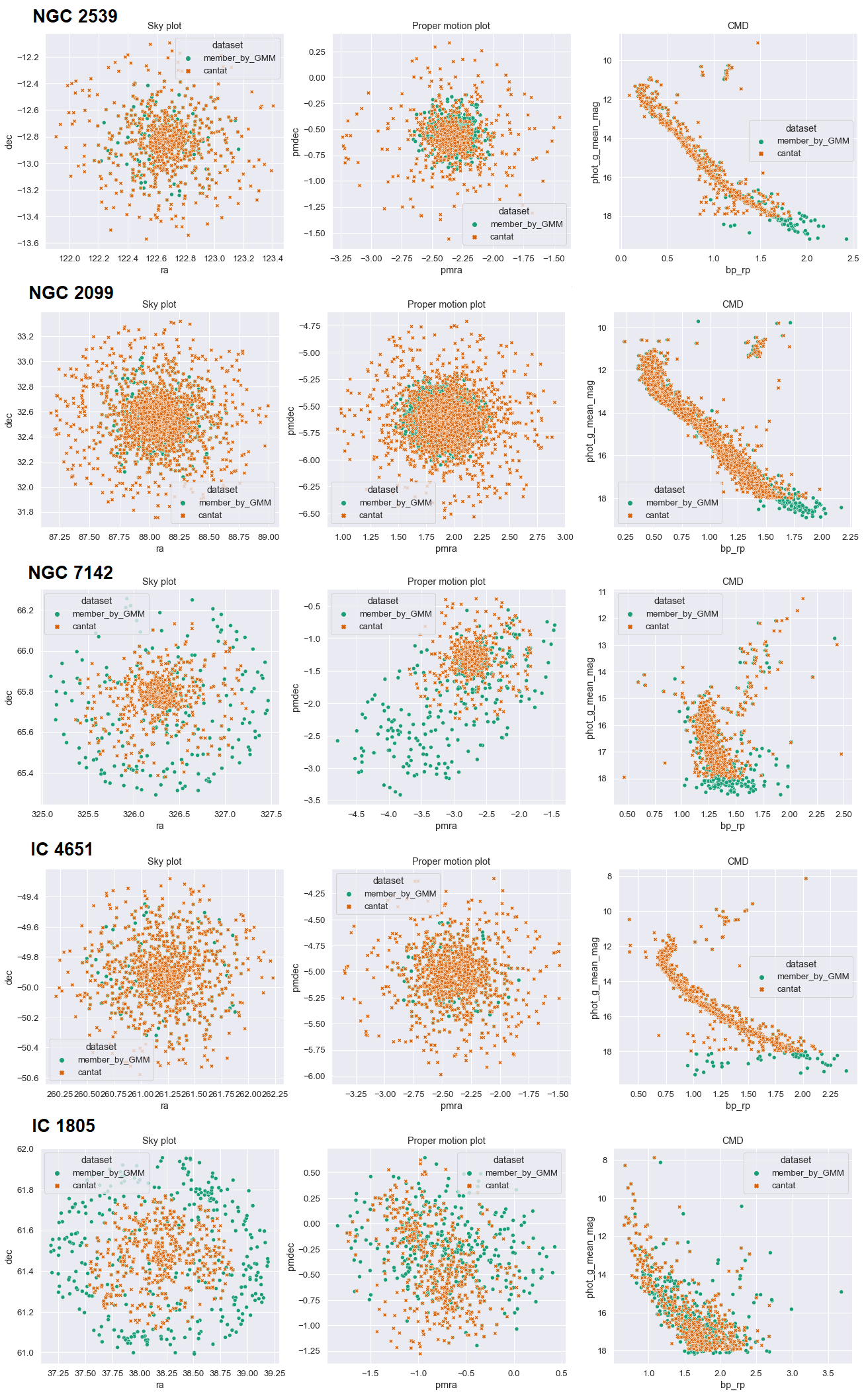}
\caption[Comparison of members]{Comparison of members defined by our method (green) and \cite{cantat2020clusters} (orange). The plots show pmra and pmdec in mas/yr, ra  and dec in degrees, $g_{mag}$ and bp-rp in mag.}
\label{cg2}
\end{figure}

\begin{figure}[h]
\centering  
\includegraphics[width=\columnwidth]{ 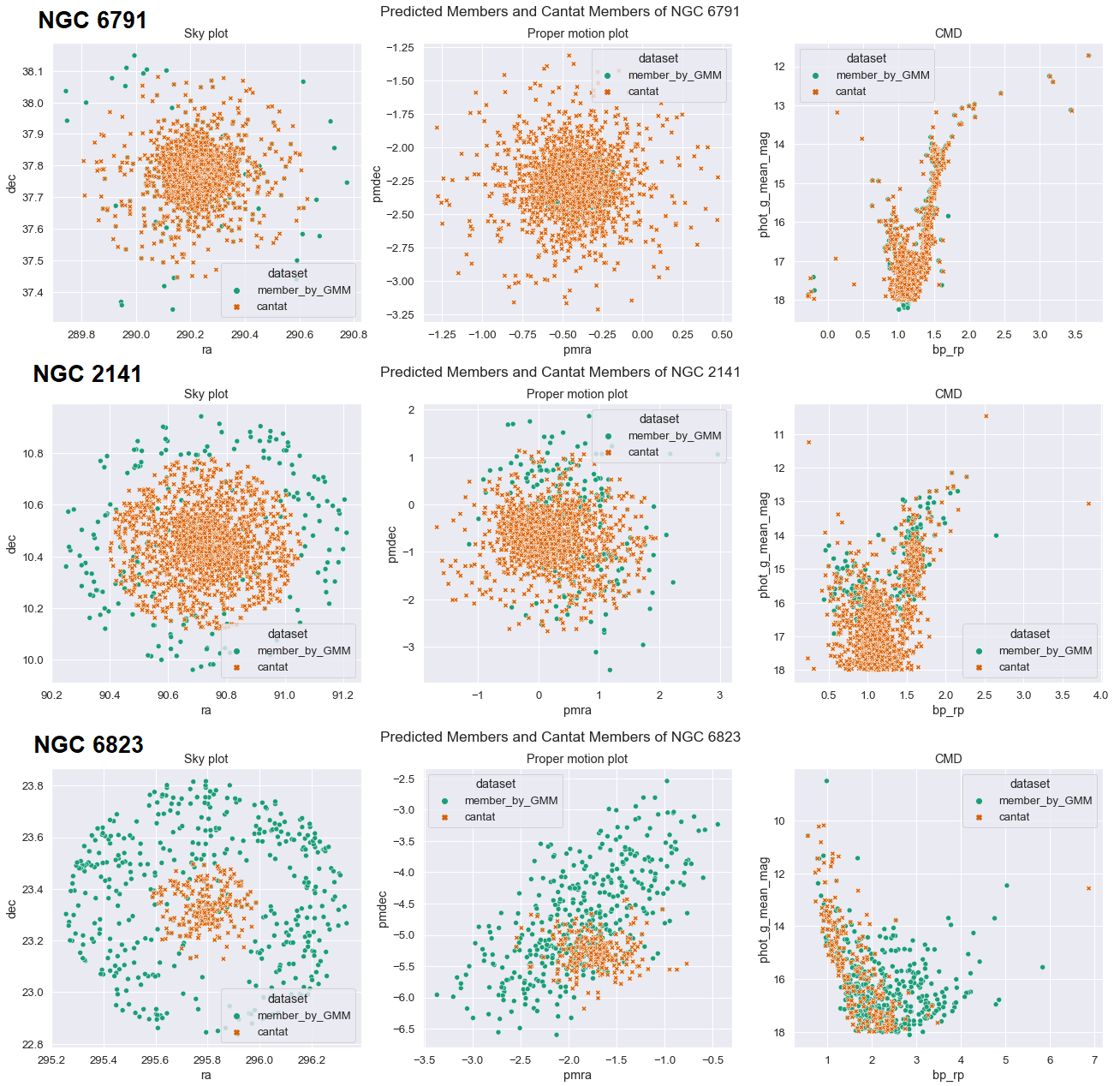}
\caption[Comparison of members]{Comparison of members defined by our method (green) and \cite{cantat2020clusters} (orange). The plots show pmra and pmdec in mas/yr, ra  and dec in degrees, $g_{mag}$ and bp-rp in mag.}
 \label{cg3}
\end{figure}

\cite{cantat2020clusters} determined members of all open clusters using the GAIA DR2. We use it as a benchmark and find fainter members at the low mass end $G \approx 20$ . As GAIA DR3 provides more precise astrometric data, it is possible that we can find a member that was earlier classified as a field star due to a lack of precise measurement. We compare the members identified by the \citep{cantat2020clusters} and our analysis, which is shown in Figs. \ref{cg1}, \ref{cg2} and \ref{cg3}.

As seen in the Figures, the proper motion scatter is large in the case of NGC~6823  and NGC 1893. All these clusters are at distances $>$
2000pc. But there are other clusters at similar distances for which the scatter is less. In the case of these two clusters we could not use APOGEE or GALAH data for validation and hence we cannot confirm our result. However, as both these clusters are young, we have found a large number of pre-main sequence stars, which are probable members. In general, caution should be exercised in the use GMM for clusters at further distances $> 3$  kpc. Supplementary methods of validation may be used in such cases.

\section{Spectroscopic Data: APOGEE and GALAH}
\begin{figure}[h]
\begin{center}
\includegraphics[width=0.6\textwidth]{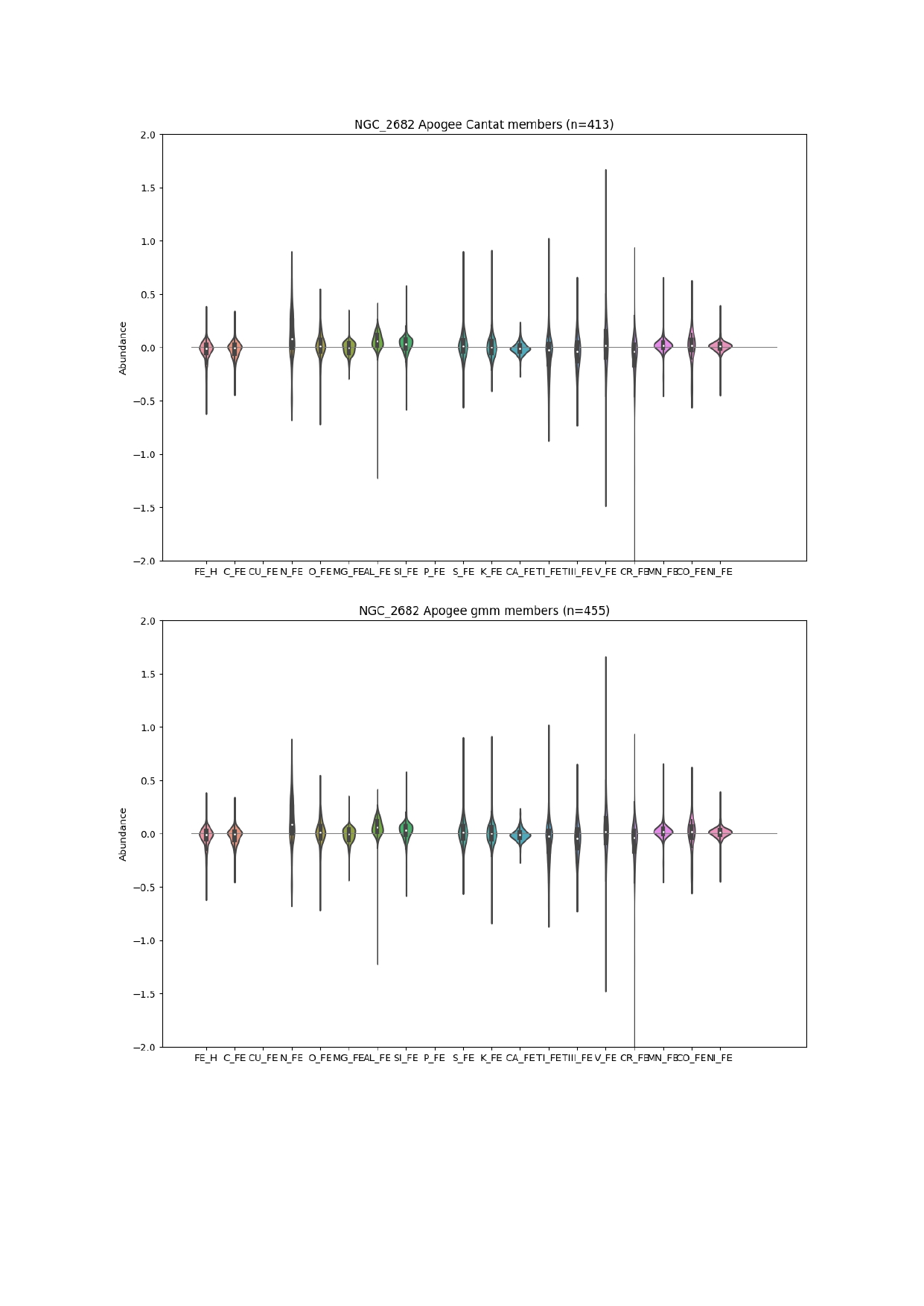}
\caption{Chemical abundances of members from APOGEE for NGC~2682 (a) Upper plot \citep{cantat2018gaia} (b) Our results.}
\label{APGM2682}
\end{center}
\end{figure}

\begin{figure}[h]
\begin{center}
\includegraphics[width=0.6\textwidth]{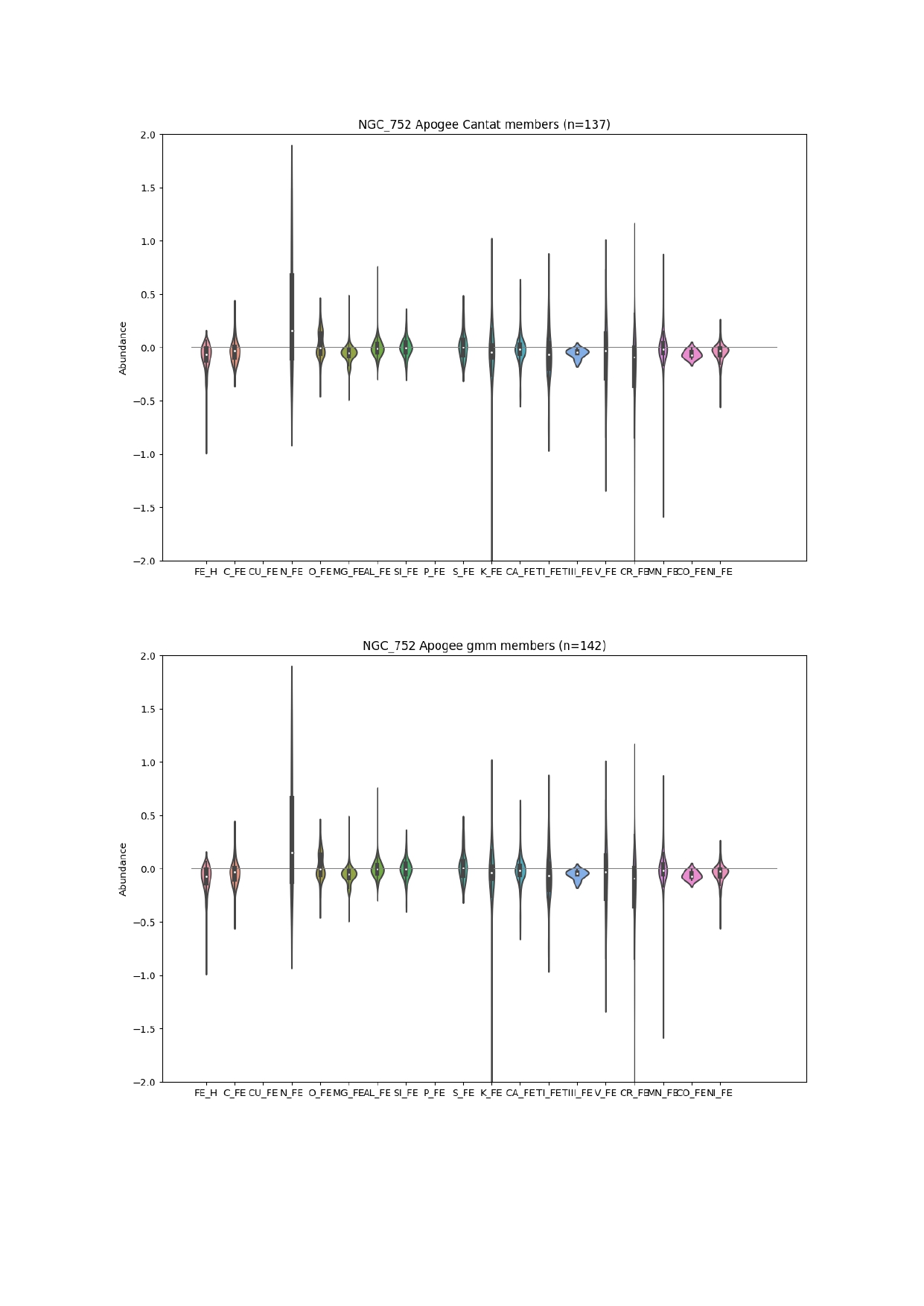}
\caption{Chemical abundances of members from APOGEE for NGC~752 (a) Upper plot \citep{cantat2018gaia} \\(b) Our results.}
\label{APGM752}
\end{center}
\end{figure}

\begin{figure}[h]
\begin{center}
\includegraphics[width=0.6\textwidth]{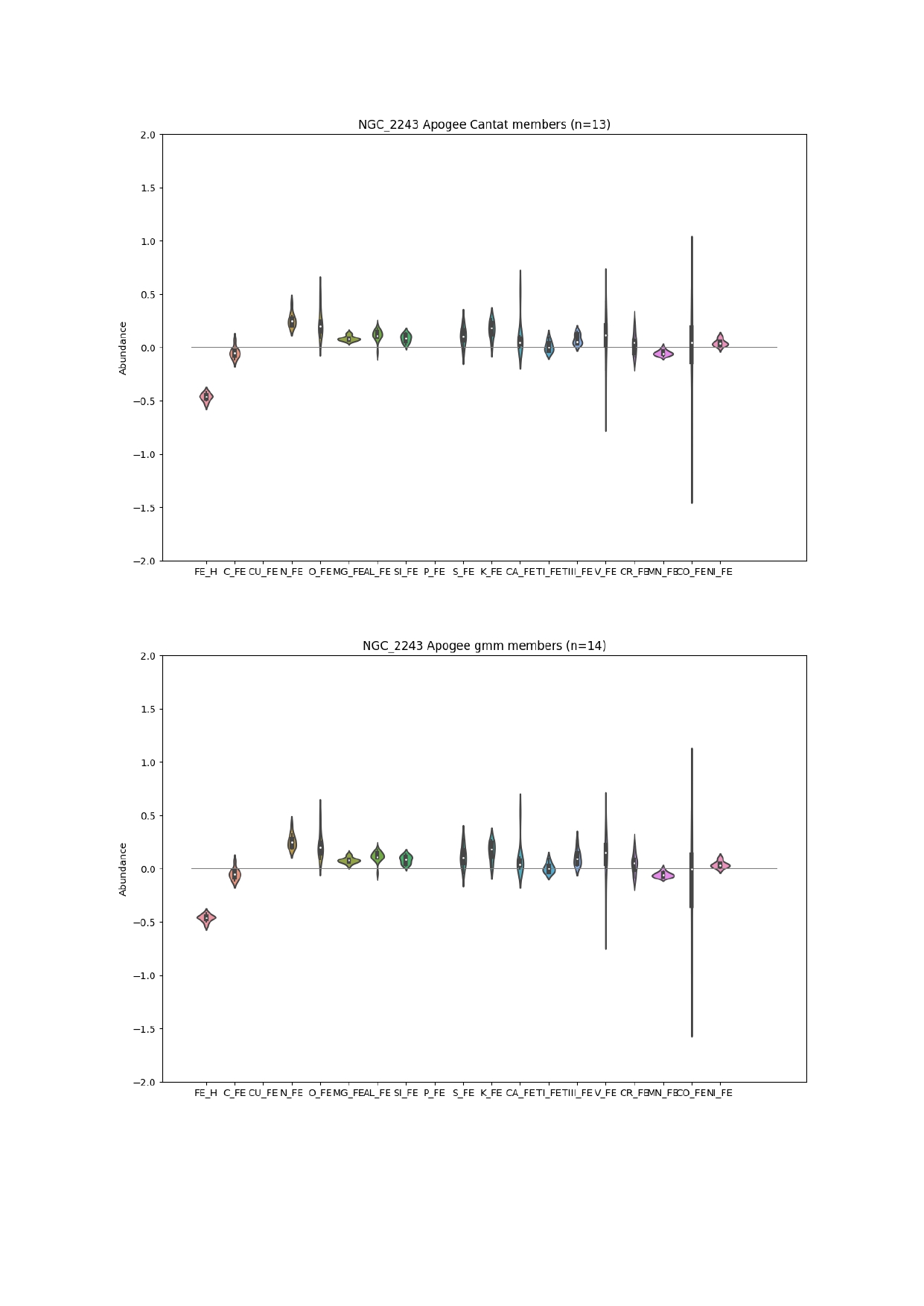}
\caption{Chemical abundances of members from APOGEE for NGC~2243 (a) Upper plot \citep{cantat2018gaia} \\ (b) Our results.}
\label{APGM2243}
\end{center}
\end{figure}

\begin{figure}[h]
\begin{center}
\includegraphics[width=0.6\textwidth]{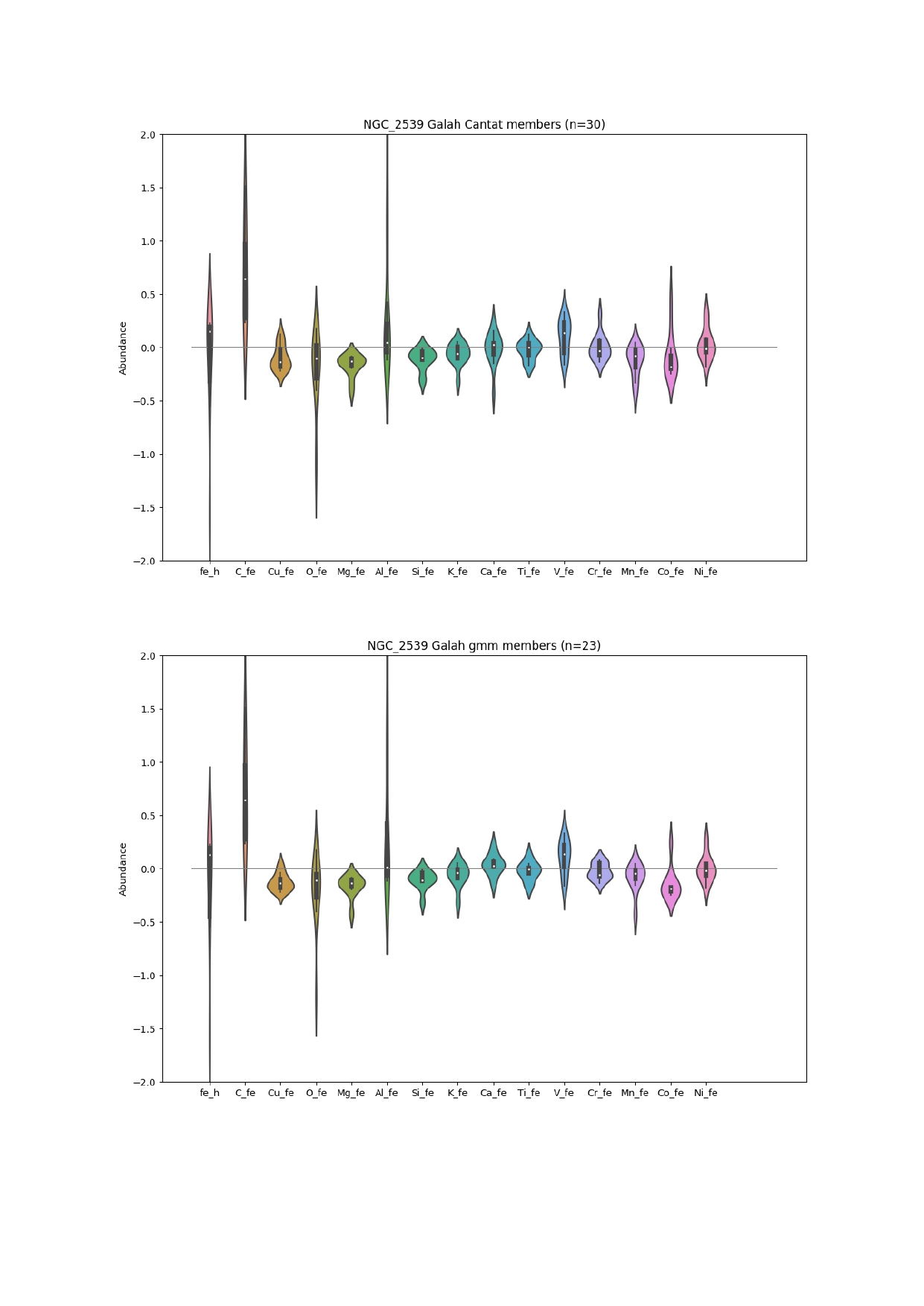}
\caption{Chemical abundances of members from GALAH for NGC~2539 (a) Upper plot \citep{cantat2018gaia} (b) Our results.}
\label{GAGM2539}
\end{center}
\end{figure}

We compared the chemical abundances of our members and the members  found by \cite{cantat2018gaia} using APOGEE and GALAH data, where available. 
APOGEE an acronym of, Apache Point Observatory Galactic Evolution Experiment,
is a large scale, stellar spectroscopic survey which is conducted in the near infra-red (IR) region of the electromagnetic spectrum.
APOGEE \citep{majewski2017apache} observations provide $R \sim 22,500 $ spectra in the infrared H-band, $1.5-1.7\mu m$, as part of the third and fourth phases of the Sloan Digital Sky Survey \citep{eisenstein2011sdss, blanton2017sloan}.
Figures \ref{APGM2682}  to \ref{APGM2243} show the chemical abundances of members from APOGEE. The upper plot shows members found by \cite{cantat2018gaia} and the lower one is our result.
The Galactic Archaeology with HERMES (GALAH) is a high resolution, ground-based spectroscopic survey.
It is carried out using the Anglo-Australian Telescope's Two Degree Field (2dF) of view and the High Efficiency and Resolution Multi-Element Spectrograph (HERMES) \citep{barden2010hermes, heijmans2012integrating, sheinis2015first}.The HERMES spectrograph gives a high resolution (R $\sim$ 28000) spectra for 392 stars in four passbands.
Figures \ref{GAGM2539} show the chemical abundances of members from GALAH. The upper plot shows members found by \citep{cantat2018gaia} and the lower one is our result.

\section{ASteCA Results}

\begin{figure}[h]
\begin{center}
\includegraphics[width=0.8\textwidth]{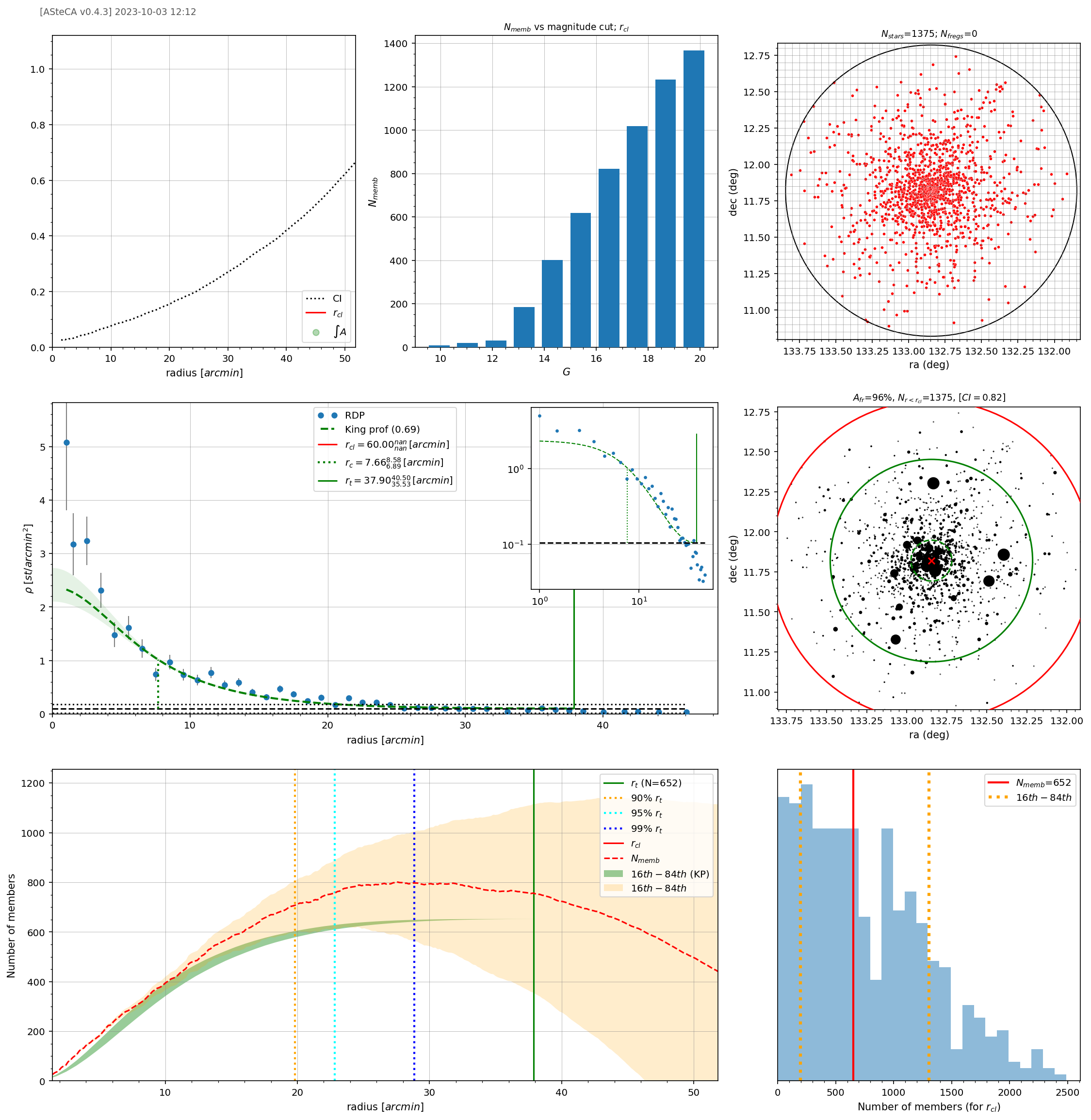}
\caption{ASteCA plots of NGC~2682}
\label{ASGM2682}
\end{center}
\end{figure}

\begin{figure}[h]
\begin{center}
\includegraphics[width=0.8\textwidth]{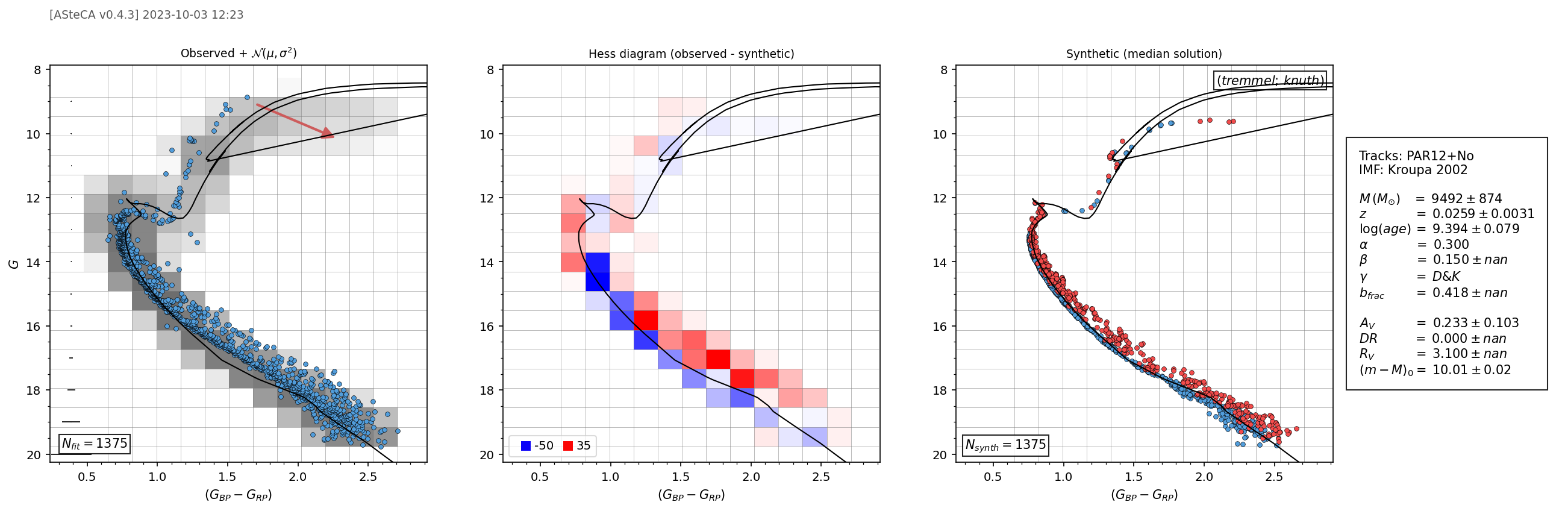}
\caption{ASteCA CMD plots of NGC~2682}
\label{ASGM2682_2}
\end{center}
\end{figure}

\begin{figure}[h]
\begin{center}
\includegraphics[width=0.8\textwidth]{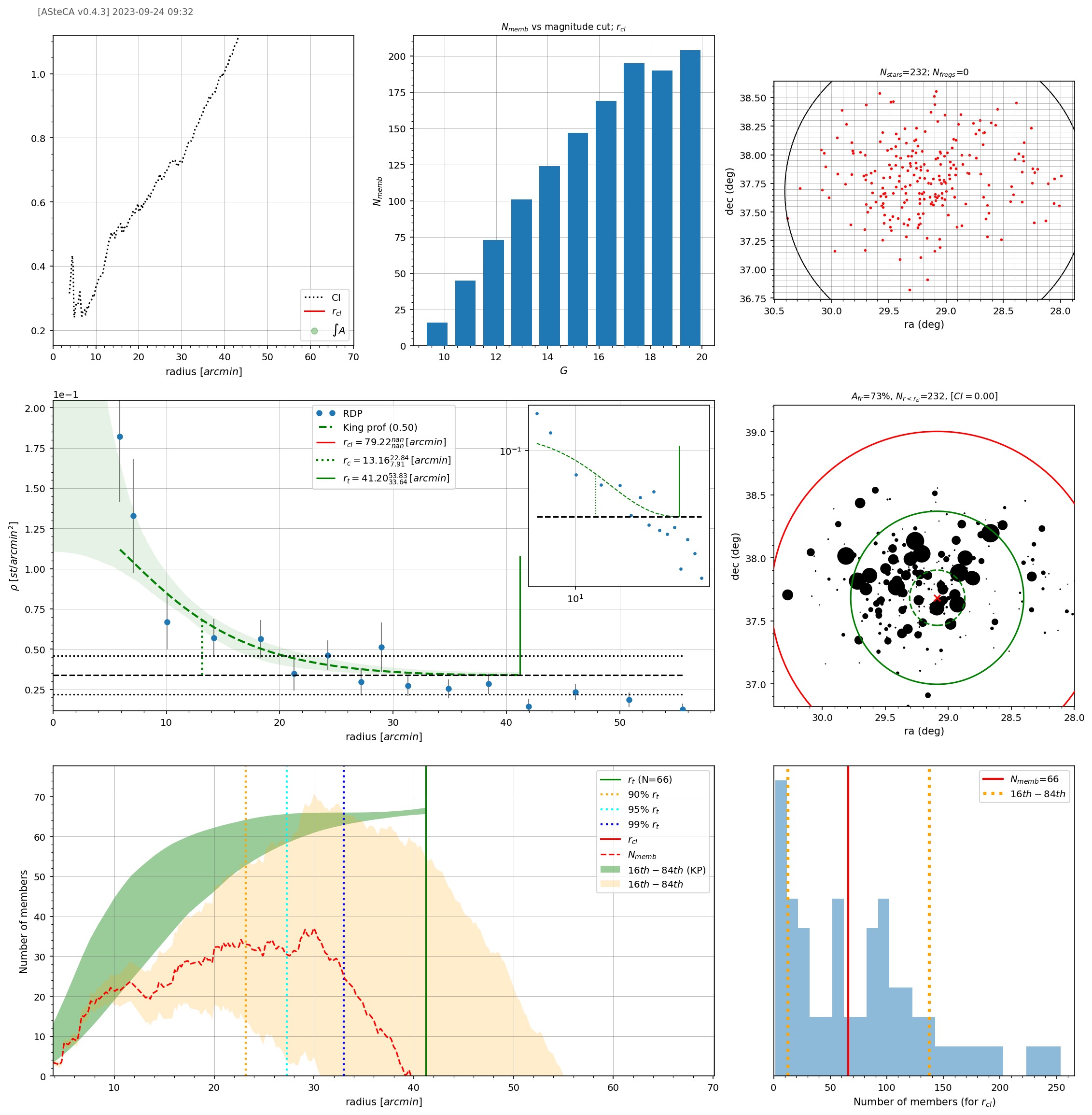}
\caption{ASteCA plots of NGC~752}
\label{ASGM752}
\end{center}
\end{figure}

\begin{figure}[h]
\begin{center}
\includegraphics[width=0.8\textwidth]{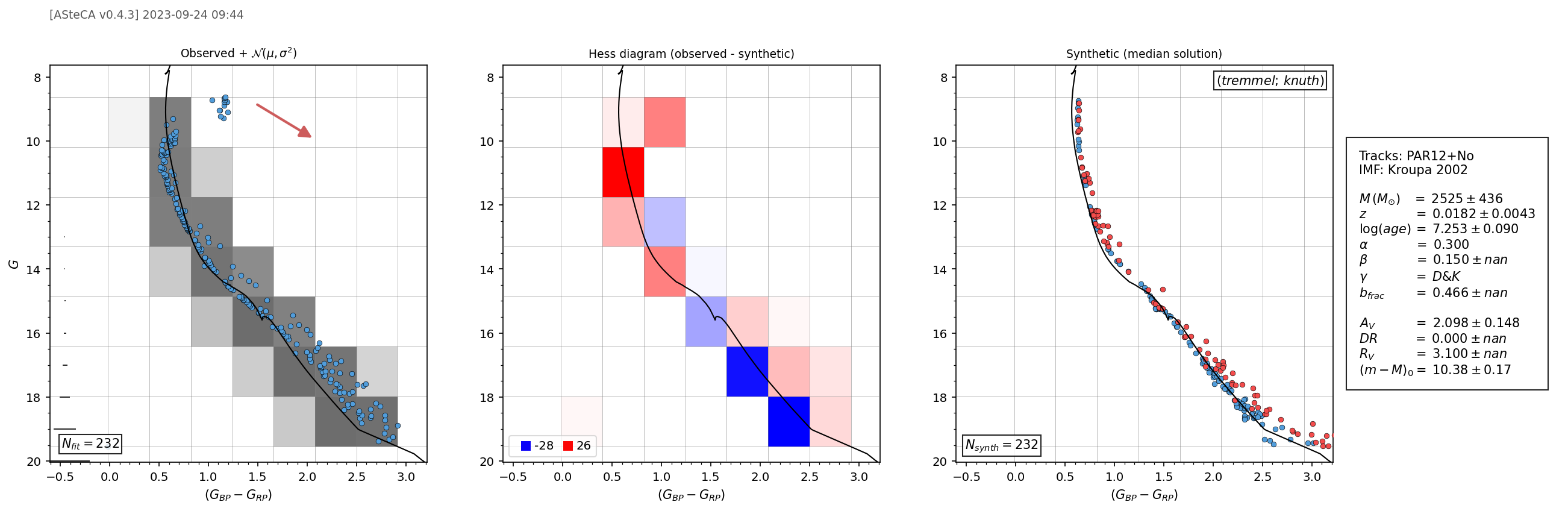}
\caption{ASteCA CMD plots of NGC~752}
\label{ASGM752_2}
\end{center}
\end{figure}

Table \ref{Astecaparamgmm} shows 
the ASteCA parameters \cite{perren2015asteca} of some of our sample clusters and see they compare well with the values of \cite{cantat2020clusters}.
Figures \ref{ASGM2682} to \ref{ASGM752_2} show the ASteCA plots and CMDs of the members obtained after running GMM clustering algorithm on our sample. 
\begin{table}
\begin{center}   
\small

\begin{tabular}{lccccc}
\hline
 & & ASteCA &  & Cantat&  \\
 
Cluster & $R_{cl}$ (arc min) & log(age) & d  & log(age) & d \\
\hline
\hline
\\
\vspace{0.5cm}
NGC~2682 & 60 &  9.39 & 913  & 9.63 & 889\\
\vspace{0.5cm}
NGC~2539 & 30.77 & 8.92  & 1137  & 8.84 & 1228 \\ 
\vspace{0.5cm}
NGC~2099 & 35.22 & 8.95 & 1282 & 8.65 & 1432\\
\vspace{0.5cm}
NGC~581 & 60.33 & 8.39 & 2333 & 7.44 & 2502\\
\vspace{0.5cm}
NGC~2243 & 29.74 & 9.436 & 4246 & 9.64 & 3719\\
\vspace{0.5cm}
NGC~7142 & 72.99 & 9.99  & 1828 & 9.49 & 2406\\

\hline
 \end{tabular}
\caption[ASteCA parameters]{ASteCA Parameters vs parameters from \citep{cantat2020clusters} using GMM.}
\label{Astecaparamgmm}
\end{center}
\end{table}

\section{Results and Discussion}

In this paper, we apply the unsupervised Gaussian Mixture Model (GMM) to a sample of thirteen clusters with varying ages ($log\ t \approx 6.38-9.64$) and distances (441-5183 pc) from Gaia DR3 data to determine
membership. \cite{cantat2018gaia}  found members to a large sample of cluster down to G $\approx$ 18 mag. We go deeper, and find members with  G $\approx$ 20 mag. We also define a quantifiable
metric Modified Silhouette Score (MSS) to evaluate its performance. We study the dependence of MSS on age, distance, extinction,
galactic latitude and longitude, and other parameters to find the particular cases when GMM seems to be more efficient than other
methods.
We find that the quality of GMM model depends on choosing an optimal filter/cutoff for the features. In our analysis, we derived this optimal cutoff for parallax and proper motion empirically where we get good $MSS$ values. We also used a combination of $k$-dist, $MNN$ and $MSS$ to determine suitable $\epsilon$ and $MinPts$ values for each cluster. We describe the improvement in the number of members and show plots where the cluster stars show peaks in parallax and proper motions with small $\sigma$ compared to the field stars. We use this feature to define our metric $MSS$ for the clustering method.

We compared MSS for all clusters with varying ages but we did not find any significant differences between GMM
performance for younger and older clusters. However we found a moderate correlation between GMM performance and the cluster
distance, where GMM works better for closer clusters as the the errors in $pmra$, $pmdec$ and parallax will increase for distant clusters. We find that GMM does not work very well for clusters at distances larger
than 3 kpc. It also depends on other factors mentioned above as well as  the field star contamination which varies with the position of the cluster in the galaxy. We use ASteca to determine parameters for the clusters from our revised membership data.

Table \ref{GMMresults} shows the results we obtained for our sample using GMM which shows the increase in number of members. Figures \ref{m67} to \ref{n581} show a variety of interesting features in the CMDs of the clusters such as pre-main sequence stars in NGC~6823, blue stragglers in M~67, NGC~2243 and NGC~7142, binaries in NGC~752, IC~4651 \& NGC~2539, gaps in NGC~581 and we obtain photometric outliers in these clusters, as photometric data was not used in the GMM model we used. 
\begin{table}
    \centering
\begin{tabular}{lrrrr}
\hline
  Cluster &  MSS &  Member &  Member & Ratio\\
    &   & GMM & Cantat & GMM/Cantat\\
\hline
\hline
\vspace{0.5cm}
  NGC 2682  & 0.94 &   1390 &       691 & 2.01\\
 \vspace{0.5cm}
 NGC 752  & 0.93 &    232 &       240 & 0.97\\
  \vspace{0.5cm}
  IC 4651  & 0.90 &    875 &       854 & 1.02\\
  \vspace{0.5cm}
 NGC 2539  & 0.90 &    560 &       518 & 0.93\\
 \vspace{0.5cm}
 NGC 2099  & 0.90 &   1607 &      1710 & 0.94\\
 \vspace{0.5cm}
  NGC 581  & 0.87 &    458 &       152 & 3.01\\
 \vspace{0.5cm}
 NGC 6823  & 0.84 &    397 &       158 & 2.51\\
 \vspace{0.5cm}
 NGC 2243  & 0.84 &    484 &       515 & 0.94\\
\vspace{0.5cm}
 IC 1805  & 0.81 &    495 &       136 & 3.63\\
 \vspace{0.5cm}
 NGC 7142  & 0.79 &    430 &       401 & 1.07\\
  \vspace{0.5cm}
NGC 6791  & 0.79 &    1106 &      1654 & 0.67\\
\vspace{0.5cm}
NGC 2141   & 0.59 &    284 &       831 & 0.34\\
 \vspace{0.5cm}
NGC 1893  & 0.51 &    592 &       169 & 3.50\\
\hline
\end{tabular}

\caption[GMM Performance]{GMM Performance for the cluster sample}
\label{GMMresults}
\end{table}

\begin{figure}[ht]
\centering  
\includegraphics[width=0.6\columnwidth]{ 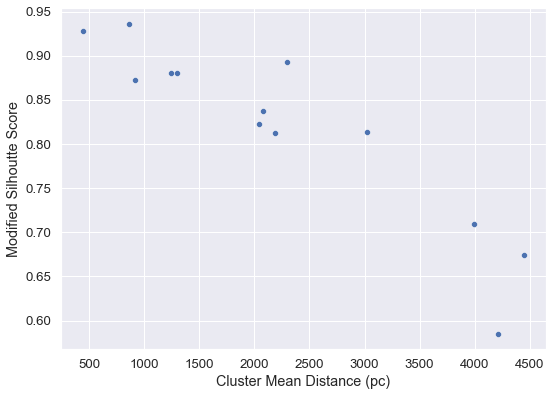}
\caption{MSS metric as a function of cluster distance. We can see a general trend that the MSS decreases (which means the model performs worse) for the distant clusters.}
\label{fig:mss_vs_dist}
\end{figure}

As noted in Table \ref{GMMresults}, the lowest MSS results are for clusters at larger distances.  The most likely parameter in these is the distance. GMM appears to work better for closer clusters and not very well for clusters at larger distances ($ > 3000$ pc). This relationship looks more clear in Fig \ref{mss_hw_clusters}. 

This is because GMM is very sensitive to field star contamination. In the case of distant clusters, foreground field stars will be brighter, more in number and dominate the sample compared to the cluster stars. In the case of closer clusters, the field stars will be fainter and contribute lesser to the sample. The accuracy depends upon the sample composition i.e., the ratio of member to non-members. If, for example, the sample consists of 90\% members of member stars and 10\% members of field stars. Then the model makes predictions with an accuracy of 90\% by correctly predicting all of the training samples that belong to member stars. If we test the same model using a test set that contains 60\% of examples from member stars and 40\% from field stars. The accuracy will then drop, and we will end up with a score of 60\%. At larger distances, the errors in $pmra$, $pmdec$ and parallax will increase and therefore GMM will work best for closer clusters. 

If we put an MSS score cut-off of $0.8$, then we should limit our cluster distance to be $< 3$ kpc. Caution should be exercised in the use GMM for clusters at further distances $> 3$  kpc. Supplementary methods of validation may be used in such cases.

\section{Acknowledgements}
This work has made use of data from the European Space Agency (ESA) mission
{\it Gaia} (\url{https://www.cosmos.esa.int/gaia}), processed by the {\it Gaia}
Data Processing and Analysis Consortium (DPAC,
\url{https://www.cosmos.esa.int/web/gaia/dpac/consortium}). Funding for the DPAC
 has been provided by national institutions, in particular the institutions
 participating in the {\it Gaia} Multilateral Agreement.





\bibliographystyle{elsarticle-harv} 
\bibliography{example}

\end{document}